\documentstyle[12pt]{article}
\newcommand{\beq}{\begin{equation}}
\newcommand{\eeq}{\end{equation}}
\newcommand{\bea}{\begin{eqnarray}}
\newcommand{\eea}{\end{eqnarray}}

\newcommand{\k}{\kappa}
\newcommand{\e}{\epsilon}

\newcommand{\reh}{\sqrt{\epsilon \hbar}}
\newcommand{\rg}{\sqrt{g}}
\newcommand{\p}{\phi}
\newcommand{\q}{\dot{q}}

\renewcommand{\d}{\delta}

\renewcommand{\L}{\Lambda}
\renewcommand{\b}{\beta}
\renewcommand{\a}{\alpha}
\newcommand{\E}{{\cal E}}
\newcommand{\G}{{\cal G}}
\newcommand{\V}{{\bf V}}

\newcommand{\n}{\nu}
\newcommand{\m}{\mu}

\newcommand{\s}{\sigma}
\newcommand{\D}{\Delta}

\newcommand{\vph}{\varphi}
\newcommand{\oh}{\frac{1}{2}}

\newcommand{\non}{\nonumber}
\newcommand{\vp}{\vec{p}}
\renewcommand{\t}{\tau}
\newcommand{\rf}[1]{(\ref{#1})}
\newcommand{\ra}{\rightarrow}

\begin{document}

\addtolength{\baselineskip}{0.20\baselineskip}
\hfill NBI-HE-93-54

\hfill gr-qc/9309010

\hfill August 1993
\begin{center}

\vspace{36pt}

{ {\Large \bf The Dispersion of Newton's Constant:} \\
  {\large \bf A Transfer Matrix Formulation of Quantum Gravity} }
\footnote{Supported by the U.S. Department of Energy under Grant
No. DE-FG03-92ER40711.}

\end{center}

\vspace{24pt}

\begin{center}
{\sl J. Greensite}
\footnote{Permanent address: Physics and Astronomy Dept., San Francisco
State University, San Francisco CA 94132.}
\footnote{Email: greensit@stars.sfsu.edu, greensite@nbivax.nbi.dk} \\

\vspace{12pt}

The Niels Bohr Institute \\
Blegdamsvej 17 \\
DK-2100 Copenhagen \O ~~Denmark \\

\vfill
{\bf Abstract}
\vspace{12pt}
\end{center}

  A transfer matrix formalism applicable to certain reparametrization
invariant theories, including quantum gravity, is proposed.  In this
formulation it is found that every stationary state in quantum gravity
satisfies a Wheeler-DeWitt equation, but each with a different value of
the Planck mass; the value $m_{Planck}^4$ turns out to be proportional
to the eigenvalue of the evolution operator.  As a consequence, the fact
that the Universe is non-stationary implies that it is not in an
eigenstate of Newton's constant.

\vfill

\newpage

\section{Introduction}

    In non-parametrized Hamiltonian mechanics, there is a set of conjugate
variables $\{q^i,p_i\}$ and a time parameter $t$. Classical trajectories
are fixed by specifying initial $\{q^i_0\}$ and final $\{q^i_f\}$
coordinates, and also a time lapse $\D t$.  In parametrized mechanics
there is also a set of conjugate variables $\{q^\m,p_\m\}$ and an
evolution parameter $\t$.  The difference is that in parametrized
theories, a classical trajectory is fixed by the initial $\{q^\m_0\}$
and final $\{q^\m_f\}$ coordinates alone; the values of $\t$
that happen to be associated with those coordinates are irrelevant.
An example is the case of a free relativistic particle, where
specifying $\{x^\m\}$ at the initial and final points determines
the trajectory.  A field-theory example is general relativity, where
the initial and final three-manifold is sufficient, together
with Einstein's equations, to determine the four-manifold between them.
Since the $\{q^\m\}$ of parametrized theories contain time information,
applying standard quantization prescriptions is, in effect, "quantizing
time".  This poses no problem when the Hamiltonian of the parametrized
theory is parabolic, as in parametrized non-relativistic mechanics, or
parametrized relativistic scalar field theory.  But for parametrized
theories with hyperbolic Hamiltonians, such as the free relativistic
particle, or quantum gravity, standard quantization procedures can lead
to serious difficulties in identifying an appropriate evolution parameter,
and a conserved non-negative norm \cite{Kuchar1}.  In some cases, e.g. a
free relativistic particle in flat space, these problems can be
easily overcome; in others, such a free relativistic particle moving in
an arbitrary curved background, they are much more problematic.  In quantum
gravity, these difficulties are known as the "problem of time."

   In this article I will propose a transfer matrix formalism for quantum
gravity, and certain other parametrized theories of the type described
above.  Since a transfer matrix, by definition, is an evolution operator,
this proposal is intended as a possible resolution of the time problem in
quantum gravity.  The method treats all dynamical degrees of freedom
of parametrized theories on an equal footing: all are operators,
none in particular is an evolution parameter.  This method will
first be illustrated for a one-dimensional "universe", whose action
is that of a free relativistic particle.  The formalism will then be
extended to minisuperspace-type actions, typical of quantum cosmology,
and finally (with caveats regarding operator-ordering and regularization)
to full quantum gravity.

\section{A One-dimensional Universe}

    A free relativistic particle is the simplest example of a system
with a reparametrization invariant Lagrangian; for this reason it is
often used as a "warmup" exercise \cite{Kuchar1,Polyakov}
for higher dimensional reparametrization invariant theories, such as
strings and quantum gravity.  To briefly recall some of the
familiar analogies:  The action of a relativistic particle
\footnote{signature convention $\eta = \mbox{diag}[-1,1,1,1]$.}

\beq
      S = - m_0 \int d\t \sqrt{-\eta_{\m\n} {dx^\m \over d\t}
{dx^\n \over d\t}}
\label{S_p}
\eeq
is invariant under reparametrizations $\t \ra f(\t)$; the
Einstein-Hilbert action is invariant under general coordinate
transformations $x^\m \ra x'^\m(x)$.  In phase
space, one can also write for the particle theory

\bea
    S &=& \int d\t (p_\m {dx^\m \over d\t} - NH)
\non \\
     H  &=& {1 \over 2m_0}(p^\m p_\m + m_0^2)
\eea
and this leads to the Hamiltonian constraint $H=0$, which is just
the mass-shell constraint.  The analogous steps, in the ADM
decomposition for gravity, lead to the superHamiltonian and
supermomentum constraints.  Upon quantization, the Hamiltonian
constraint becomes a constraint on physical states $H \psi = 0$,
which is the Klein-Gordon equation in the case of a particle, and
the Wheeler-DeWitt equation for quantum gravity.

    If one slices spacetime into a series of spacelike hypersurfaces
$\Sigma_T$ with a parameter $T = T(x)$ (the time coordinate) labeling
each hypersurface, the following norm is preserved by the
Klein-Gordon equation:

\beq
      <\psi|\psi>_T =  {i \over 2m_0}\int d\Sigma_T^\m (\psi^* \partial_\m
\psi - \psi \partial_\m \psi^*)
\label{norm}
\eeq
This norm is not positive definite in general (although in flat
spacetime one can restrict to positive frequency states, in which case
the norm {\it is} positive definite).\footnote{More generally, it is
possible to identify a conserved non-negative norm in spacetimes with
a timelike Killing vector field.}  A similar construction can be
made in quantum gravity.  It is possible to slice superspace into
hypersurfaces such that a norm analogous to \rf{norm} is
independent of the hypersurface, providing $\Psi$ satisfies the
Wheeler-DeWitt equation.  It is not clear how to make such a norm
non-negative.  Nevertheless, the analogy to the relativistic particle
case suggests (and the idea goes back to the classic paper of DeWitt
\cite{DeWitt}) that some coordinate in superspace can be interpreted as a
time evolution parameter in quantum gravity.  There have been many subsequent
variations and extensions of this idea; a recent comprehensive review is
found in \cite{Isham}.

   On the other hand, there is one aspect of the relativistic particle
example which seems quite different from the situation in 4D gravity.
In the case of the relativistic particle, the observer is obviously {\it
external} to the particle.  The information contained in the
Klein-Gordon wavefunction refers to measurements that can be made
by such external observers, who are free to measure the particle
position on any spacelike hypersurface.  In contrast, observers in
4D gravity (at least, human observers) are {\it internal} to the system
in question (the Universe), since we live in spacetime rather than
superspace.  Is it possible to find and quantize a one-dimensional
model which reflects also {\it this} aspect of the 4D case?

     Such a model is obtained by simply reinterpreting the action
of eq. \rf{S_p}.  Consider an observer
living in a one-dimensional universe parametrized by a single coordinate
$\t$, who is able to make observations on a 4-component "field"
$x^\m(\t)$ in that one-dimensional spacetime.  The reparametrization
invariant action of this field is taken to be that of eq. \rf{S_p}.
The Hilbert space consists of wavefunctions $\phi(x^\m)$ with norm

\beq
       <\phi|\phi> = \int d^4x \m(x) \phi^*(x^\m) \phi(x^\m)
\label{norm1}
\eeq
where $\m(x)$ is a measure to be determined.  Note that since all components
$x^\m$ are observable, the integral is taken over the full four-dimensional
"field" space.  The fact that all components $\{x^\m \}$ are observable
and all are integrated over in the norm \rf{norm1} is the main (and
crucial) difference between the "one-dimensional universe", in which
none of the observables is to be interpreted as an evolution parameter,
and the relativistic particle.

   In classical relativistic mechanics, the dynamics of a particle
moving in curved space, or in some external potential, can be
described covariantly by a parametrized trajectory $x^\m(\t)$.  In
quantum physics the configuration space is Hilbert space, and dynamics
can be presumably also be described as a parametrized trajectory
$\psi(x^\m,\t)$ in Hilbert space.  Let us postulate a corresponding
Schrodinger equation

\beq
       i\hbar \partial_\t \psi(x,\t) = H \psi(x,\t)
\label{S_eq}
\eeq
where $H$ is an operator, hermitian in the measure $\m(x)$ and invariant
w.r.t Lorentz transformations, such that
\beq
        x^\m_{cl}(\t) \equiv <\psi(x,\t)|x^\m|\psi(x,\t)>
\label{xcl}
\eeq
is a parametrized solution of the classical equations of motion.
In this way, the parameter $\t$ running
along a trajectory of Hilbert space can be identified with the variable $\t$
parametrizing a certain classical trajectory, corresponding to the
motion of the center of the wavepacket.  Of course
reparametrizations of $\t$ have no physical significance, and
the Schrodinger equation above can be made to look
covariant w.r.t reparametrizations of $\t$ by intoducing an "einbein"
for the trajectory
\beq
       i\hbar \partial_\t \psi(x,\t) = e(\t) H \psi(x,\t)
\label{Sr}
\eeq
However, assuming that $H$ is $\t$-independent, the trajectory through Hilbert
space depends only on an initial state, and not on $e(\t)$, which can
always be set to $e=1$ by a reparametrization.  From here on we
set $e=1$.

   I will now show how an evolution operator $H$ and measure $\m(x)$, with the
required properties, may be obtained from a transfer matrix formalism.
The starting point is the Euclidean action corresponding to eq. \rf{S_p},
obtained rotating the field space metric ($\eta_{\m\n}$) from Lorentzian
to Euclidean signature, and extracting an overall factor of $i$; i.e.
\beq
     iS_{Lorentz}[g_{\m\n}=\eta_{\m\n}] \ra
          -S_{Euclid}[g_{\m\n}=\d_{\m\n}]
\eeq
Then, generalizing from $D=4$ to arbitrary $D$, the transfer matrix
${\cal T}_\e$ is defined by

\bea
      \psi(y,\t+\e) &=& {\cal T}_\e \psi(y,\t) =
\exp[-H_\e \e /\hbar] \psi(y,\t)
\non \\
           &=& \int d^Dx \; \m(x) \exp(-S[y,x]/\reh) \psi(x,\t)
\label{tmat}
\eea
where $S[y,x]$ is the Euclidean action of the classical trajectory connecting
(and terminating at) points $x^\m$ and $y^\m$.  Since $S[x,y]=S[y,x]$,
and $S[y,x]$ is real, it follows that $H_\e$ is hermitian in the
measure $\m$.  Note that by the usual trick of integration by parts,
we have

\bea
      0 &=& \int \left( \prod_n d^Dx \; \m(x_n) \right)
\left\{ {1 \over \reh} {\partial S[\{x_i\}] \over \partial x^\m_k} -
{\partial \over \partial x^\m_k} \ln\m(x_k) \right\}
     \exp\left[- S[\{x_i\}]/\reh \right]
\non \\
      &=& <...\left\{ {1 \over \reh}{\partial S[\{x_i\}] \over \partial x^\m_k}
- {\partial \over \partial x^\m_k} \ln\m(x_k) \right\} ...>
\label{Efest}
\eea
where
\bea
     {\partial S[\{x_i\}] \over \partial x^\m_k} &=& {\partial \over
\partial x^\m_k} \sum_n S[x_{n+1},x_n]
\non \\
       &\ra& {\d S[x(\t)] \over \d x^\m(\t')}
\eea
and where $S[x(\t)]$ is the continuum action of the path $x^\m(\t)$.
Formally then, in the $\e \ra 0$ limit, we obtain
the quantum version of the Euclidean equations of motion
\beq
<...\d S / \d x...>=0
\eeq
The measure is given by
\beq
      \m^{-1}(y) = (\reh )^D \lim_{\e \ra 0}\int
{d^Dx \over (\reh )^D} \; \exp(-S[y,x]/\reh )
\label{measure}
\eeq
This expression is motivated by requiring that (i) the
transfer matrix is the identity operator in the $\e \ra 0$ limit; and
(ii) symmetries of the action become symmetries of the integration measure.
Finally, the evolution operator $H$ in \rf{S_eq} is obtained by the
$\e \ra 0$ limit
\beq
          H \equiv \left[ \lim_{\e \ra 0} H_\e \right]_{\d \ra \eta}
\label{Hlim}
\eeq
followed by a rotation back to Lorentzian signature ($\d \ra \eta$).

    This defines the quantization procedure.  Next I will compute
the operator $H$ and obtain the classical limit of \rf{S_eq}, first with
a flat space metric $g_{\m\n}=\eta_{\m\n}$, then with a curved space
metric.

   Let $z^\m=x^\m-y^\m$.  Since a rotation to Euclidean signature has
been performed, we can easily evaluate the integral of equation
\rf{measure}

\bea
        S[y,x] &=&  m_0\sqrt{\d_{\m\n} z^\m z^\n} = m_0 |z|
\non \\
        \m^{-1}(y) &=& \int d^Dz \exp[-{m_0 |z| \over \reh }] =
{2 \pi^{D/2} \over \Gamma(D/2)} (D-1)! ({\reh \over m_0} )^D
\eea
and eq. \rf{tmat}

\bea
       \psi(y,\t+\e) &=& \int d^Dz \; \m \exp[-{m_0 |z| \over \reh }]
\left\{ \psi(y) + {\partial \psi \over \partial y^\m}z^\m \right.
\non \\
 & & \left. + \oh{\partial^2 \psi \over \partial y^\m \partial y^\n}z^\m z^\n
+ O(z^3) \right\}
\non \\
   &=& \left[ 1 + \e \hbar {D+1 \over 2m_0^2}
\d^{\m\n} \partial_\m \partial_\n + O(\e^2)\right] \psi(y,\t)
\eea
Identifying $H_\e$, and taking the limit and signature rotation
prescribed in \rf{Hlim}, we obtain

\beq
  H = - \hbar^2 {D+1 \over 2m_0^2} \eta^{\m\n} \partial_\m \partial_\n
\eeq
and it is easy to see that the eigenstates of $H$

\beq
         \psi_\E (x^\m,\t) = e^{-i\E \t /\hbar} \p_\E (x^\m)
\eeq
are all solutions of the Klein-Gordon equation
\beq
  (\hbar^2 \eta^{\m\n} \partial_\m \partial_\n +
          {2m_0^2 \E \over D+1}) \p_\E = 0
\eeq
where the mass-shell parameter
\beq
   m^2 = -{2m_0^2 \E \over D+1}
\eeq
is proportional to
the eigenvalue of the evolution operator $H$.

   The important point to notice, in this very simple example, is that
the transfer matrix formalism has not destroyed the constraint, since
each stationary state satisfies the usual Klein-Gordon equation.  What
is unusual is that the dimensionful parameter in the constraint, in
this case $m^2$, is different for each stationary state.  In effect,
$m^2$ has become a dynamical quantity, like energy, rather than a fixed
parameter.

   The classical equations of motion can be obtained by WKB methods, or
more simply by just making the replacement

\bea
       H[x^\m,-i\hbar \partial_\m] &\ra& H[x^\m,p_\m]
\non \\
         H_c[x,p] &=& \lim_{\hbar \ra 0} H[x,p]
\label{Hclass}
\eea
and applying the classical equations
\beq
  H_c=\E,~~~~~~~~{dx^\m \over d \t} = {\partial H_c \over \partial p_\m},
       ~~~~~~~~{dp_\m \over d \t} = -{\partial H_c \over \partial x^\m}
\eeq
where $\E$ is fixed from the initial conditions.  This gives us

\beq
  \eta^{\m\n}p_\m p_\n = -M^2,~~~~~~~~~{dx^\m \over d\t}
= {D+1\over m_0^2} \eta^{\m\n} p_\n
\eeq
where $M^2= -2m_0^2 \E /(D+1)$.  This is equivalent, of course, to the
usual classical equation of motion for a free relativistic particle
moving in flat space, which is obtained by the rescaling

\beq
          \t = {m_0^2 s \over M (D+1)}
\label{proptime}
\eeq
where $s$ is the proper time of the particle worldline.  The
proportionality of $\t$ to the proper time $s$ has no special
significance since, as noted below eq. \rf{xcl}, the parametrization
can be modified simply by choosing a different $e(\t) \ne 1$.

   The non-relativistic limit of this theory is equally trivial, but
still instructive.  Making factors of $c$ explicit, the general solution
to the evolution equation \rf{S_eq} (with $p^0>0$) is

\beq
    \psi(x^\m,\t) = \int dm \int d \vp \; f(m,\vp)
\exp\left\{i[{D+1 \over 2m_0^2} m^2 \t -\sqrt{m^2c^4 + \vp^2c^2}t +
\vp \cdot \vec{x}]/\hbar \right\}
\eeq
where $t=x^0/c$.  For simplicity, suppose that:  i) $f(m,\vp)$ factorizes
into $f(m,\vp)=g(m)h(\vp)$; ii) $g(m)$ is very sharply peaked around
some value $M$; and iii) $h(\vp)$ is negligible unless $|p| << Mc$.
In that case, again rescaling
\beq
          \t = {m_0^2 c^2  \over M (D+1)}s
\eeq
so that $s$ has units of time, the solution can be written

\beq
      \psi(x^\m,s) \approx \vph(s,t) \psi_{NR}(t,\vec{x})
\eeq
where
\bea
       \vph(s,t) &=& \int dm \; g(m) \exp[i({m^2 c^2 \over 2M}s
- mc^2t)/\hbar]
\non \\
     \psi_{NR}(t,\vec{x}) &=& \int d\vp \; h(\vp)
\exp[-i({\vp^2 \over 2M}t - \vp \cdot \vec{x})/\hbar]
\eea
and, of course, $\psi_{NR}$ is a solution of the non-relativistic
Schrodinger equation for a free particle of mass $M$.

   Now if $g(m)$ and $f(\vp)$ are both, e.g., gaussians, then the
spreading of the wavepacket in both $t$ and $\vec{x}$ is given by

\bea
        \D t(s) = [\D t_0^2 + ({\D m \over M}s)^2]^{1/2}
\non \\
        \D x(t) = [\D x_0^2 + ({\D p \over M}t)^2]^{1/2}
\eea
where
\beq
        \D m \D t_0 \sim {\hbar \over c^2},~~~~~~~~~~~~~~~
        \D p \D x_0 \sim \hbar
\eeq
Since $\overline{t} \equiv <t> = s$, we can write $\D t(s) =
\D t(\overline{t})$.  Finally, it is consistent with the
$\D m \D t_0$ uncertainty relation to take both $\D m$ and
$\D t_0$ proportional to $1/c$.  This means that the dispersion
of the wavepacket in the time direction

\bea
\D t(\overline{t}) &=& [\D t_0^2 + ({\D m \over M}\overline{t})^2]^{1/2}
\non \\
                   &\sim& {1 \over c} \ra 0
\eea
goes to zero in the non-relativistic, $c \ra \infty$ limit.

   The lesson of this simple exercise is that, in the $c \ra \infty$
limit, the one-dimensional universe has acquired a clock.  By a "clock",
I mean a non-stationary observable (or set of observables) whose dispersion is
negligible, and whose evolution is independent of the other observables
in the system.  Measuring such an observable gives a value for the
evolution parameter, which itself is {\it not} an observable.  In the
one-dimensional universe, the "clock" is the t-component of the
field $x^\m(\t)$; this is because its dispersion can be made to vanish,
and its value becomes perfectly correllated with $\t$, in the
non-relativistic limit.  Equivalently, we can just say that the observable
$t$ behaves classically in the $c \ra \infty$ limit.

   The extension of the transfer matrix formalism to a 1-dimensional
universe with a curved field-space metric $g_{\m\n}$
is straightforward.  The reparametrization invariant
action in this case is

\beq
      S =  - m_0\int d\t \sqrt{-g_{\m\n} {dx^\m \over d \t}
{dx^\n \over d \t}}
\label{S_curv}
\eeq
Again we rotate from Lorentzian to Euclidean signature, extract an
$i$, and obtain the action for Riemannian metrics.
To evaluate the integrals in eq. \rf{tmat} and \rf{measure}, it is useful
to introduce Riemann normal coordinates $\xi^\m$ around the point $y^\m$.
The classical action for a trajectory running from $\xi^\m_1=0$ to
$\xi^\m_2=\xi$ is

\beq
      \D S[0,\xi] = m_0\sqrt{\d_{ab}\xi^a \xi^b + O(\xi^5)}
\eeq
The $O(\xi^5)$ terms will not contribute to $H$ in the $\e \ra 0$
limit, and can be dropped.  The measure is then

\bea
  \m^{-1}(y) &=& (\reh )^D \lim_{\e \ra 0} \int
{d^D\xi \over (\reh )^D}  \;
\det[{\partial z^\m \over \partial \xi^\n}]
     \exp[-m_0 |\xi|/\reh]
\non \\
             &=& {2 \pi^{D/2} \over \Gamma(D/2)} (D-1)! ({\reh \over
m_0})^D {1 \over \sqrt{g(y)}} = {\s \over \sqrt{g(y)}}
\eea
The transfer matrix is computed as in the flat-space case,
again with the help of Riemann normal coordinates

\bea
       \psi(y,\t+\e) &=& \int {d^4z \over \s }
 \sqrt{g(y+z)}  \exp[-m_0\sqrt{g_{\m \n}z^\m z^\n}/\reh ] \psi(y+z,\t)
\non \\
   &=&  \int {d^4\xi \over \s } (1 - {1\over 6}R_{\a \b} \xi^\a
\xi^\b + ...)
 \exp[-{m_0 |\xi| \over \reh }]
\left\{ \psi(y,\t) + {\partial \psi \over \partial \xi^\m}\xi^\m \right.
\non \\
 & & \left. + \oh{\partial^2 \psi \over \partial \xi^\m \partial \xi^\n}
\xi^\m \xi^\n + O(\xi^3) \right\}
\non \\
   &=& \left[ 1 + \e \hbar {D+1 \over 2m_0^2} \partial^\m \partial_\m -
\e \hbar {D+1 \over 6m_0^2} R + O(\e^2) \right] \psi(y(\xi),\t)
\eea
where $R$ is the curvature scalar.  Transforming back from Riemann
normal coordinates and taking the limit \rf{Hlim} gives

\beq
      H = -  {D+1 \over 2m_0^2} \hbar^2  {1 \over \sqrt{g}} \partial_\m
\sqrt{g} g^{\m \n} \partial_\n + \hbar^2 {D+1 \over 6m_0^2} R
\label{Hcurv}
\eeq
where $g=|\mbox{det}(g_{\m\n})|$.
The appearance of the curvature scalar term in the evolution operator
$H$ is related to the choice of measure, and, in turn, the ordering
of operators in eq. \rf{Hcurv}. There are other, more complicated
choices of measure that could have been made in this problem, but
these would only affect the coefficient of $R$ (c.f. ref.
\cite{Kuchar2,Halliwell}).

   The classical equations of motion are again obtained either from
the WKB approximation, or else by the prescription \rf{Hclass},
which gives
\beq
      H_c = {D+1 \over 2m_0^2} g^{\m \n} p_\m p_\n
\eeq
Setting $H_c=\E$, and $M^2 \equiv -2m_0^2\E / (D+1)$,
and also rescaling $\t$ according to \rf{proptime}, we recover the
the classical equations of a relativistic particle
moving in curved space, with the identification
of the evolution parameter $\t$ as the proper time.

  At the quantum level, eigenstates of the mass-shell parameter
are stationary states of the evolution operator $H$.  The remarks
made above for the flat-space case, noting that all stationary states
obey the constraint equation but with different values of the mass-shell
parameter, of course apply to the curved space example also.
Non-stationarity, in the one-dimensional universe, must be attributed to
dispersion in the mass-shell parameter $M^2$.

  As a final remark we may ask whether, since the evolution parameter is
identified with proper-time in the classical limit (given the choice
$e(\t)=1$ in eq. \rf{Sr}), the proper-time could
have been used as an evolution parameter from the beginning.  This might
be done by replacing $S[y,x]$ and $\e$ in eq. \rf{tmat} with a weight
$S[(y,s+\D s),(x,s)]$, representing the minimal action between points
$x$ and $y$ of paths constrained to have proper time $\D s$.  The
problem with this is that the amplitude between $(x,s_1)$ and
$(y,s_2)$ would not, in general, be dominated by the classical path,
unless the proper time difference $s_2-s_1$ happened to correspond
to the path length of the geodesic between $x$ and $y$.  The problem
can be fixed, for Green's functions, by integrating over $s_2-s_1$
(see, e.g., \cite{Teitelboim}), but then the proper-time loses its
function as an evolution parameter.  In higher dimensions the situation
is further complicated by the fact that a proper-time slicing of a
simply connected Riemannian D-manifold into \break
(D-1)-manifolds can introduce
spurious singularities.  Under proper-time evolution, a simply connected
(D-1)-manifold will in general evolve into a set
of disconnected \break
(D-1)-manifolds \cite{Vilenkin}.  It seems unlikely that
such an approach would yield a hermitian evolution operator, although
it may have other applications (c.f. \cite{Noboru}).

\section{Minisuperspace Models}

   Next we consider actions of the form

\bea
     S &=& \int dt (p_n {dq^n \over dt} - N {\cal H})
\non \\
     {\cal H} &=&  {1 \over 2 m_0} G^{nm}p_n p_m + m_0 V(q)
\label{Smini}
\eea
where $m_0$ is some dimensionful parameter, and the supermetric
$G_{nm}$ has Lorentzian $(-++...+)$ signature. Actions of this kind
typically arise in minisuperspace models of quantum cosmology.
To compute $S[q+\D q,q]$, begin with the Hamiltonian equation

\beq
       \dot{q}^n = N{\partial {\cal H} \over \partial p_n}
= {N\over m_0} G^{nm} p_m
\eeq
insert this into the constraint equation
\beq
 {\cal H} ={m_0 \over 2N^2}G_{nm} \dot{q}^n \dot{q}^m  + m_0 V = 0
\eeq
and solve for the lapse
\beq
      N = [-{1 \over 2V}G_{nm} \dot{q}^n \dot{q}^m]^{1/2}
\eeq
Then

\bea
      S[q, q+\D ] &=& \int^{\D t}_0 dt \; {m_0 \over N} G_{nm}
\dot{q}^n \dot{q}^m
\non \\
           &=& -m_0 \int dt \sqrt{-2VG_{nm} \dot{q}^n \dot{q}^m}
\non \\
           &=& -m_0 \sqrt{-2VG_{nm} \D q^n \D q^m}
\non \\
           &=&  -m_0\sqrt{-\G_{nm} \D q^n \D q^m}
\label{Sq}
\eea
where we define a modified supermetric
\beq
          \G_{nm} \equiv 2 V G_{nm}
\label{Gcal}
\eeq
{}From here, the procedure follows exactly the same steps as in
the curved metric example of the previous section.
Again, rotate the signature of the supermetric $G_{nm}$ ({\it not}
the spacetime metric $g_{\m \n}$) to Euclidean signature and extract
a factor of $i$ to obtain the "Riemannian" action. Then,
introducing Riemann normal coordinates in minisuperspace around the
point $q+\D q$, evaluate the relevant integrals for the measure
and transfer matrix.  The final result for the evolution operator
is

\beq
      H = -  {D+1 \over 2m_0^2} \hbar^2  {1 \over \sqrt{\G}} {\partial
\over \partial q^n} \sqrt{\G} \G^{nm} {\partial \over \partial q^m}
 +  \hbar^2 {D+1 \over 6m_0^2} {\cal R}
\label{Hmini}
\eeq
where, in this case, $D$ is the dimensionality of minisuperspace,
${\cal R}$ is the curvature scalar formed from the modified supermetric
\rf{Gcal}, and
\beq
           \G \equiv |\mbox{det}(\G_{mn})|
\eeq

   It should be noted that $H$ is hermitian in the
measure $\m(q)=\sqrt{\G}$.  This is despite the fact that, in quantum
cosmology models, $V(q)$ is not positive definite, and therefore
$S[q,q+\D q]$ can be imaginary, even after rotation
of $G_{nm}$.  Nevertheless, carrying out the relevant integrals and
rotating back to Lorentzian signature, one still finds that $H$ is
hermitian in the appropriate measure.  This can be understood from
the fact that, for modified supermetrics $\G_{mn}$ of Euclidean
signature, the operator $H_\e$ is hermitian by construction, since
$S[q,q']$ is real and symmetric.  But since the hermiticity of
$H_\e$ does not depend on the precise functional form of $\G_{mn}$,
the continuation to arbitrary signature can only upset hermiticity
if it introduces factors of $i$ in $H$.  These factors of $i$ are
avoided by making the standard Euclidean $\ra$ Lorentzian continuation
$\sqrt{det(g)} \ra \sqrt{-det(g)}$, in the one-dimensional Universe
example, and $\sqrt{det(\G_{mn})} \ra \sqrt{|det(\G_{mn})|}$, in the
minisuperspace case, for both $H$ and $\m$.

   The Schrodinger equation for stationary states is now

\beq
      [-{D+1 \over 2m_0^2} \hbar^2  {1 \over \sqrt{\G}} {\partial
\over \partial q^n} \sqrt{\G} \G^{nm} {\partial \over \partial q^m}
 +  \hbar^2 {D+1 \over 6m_0^2} {\cal R} ] \Psi = \E \Psi
\eeq
or, equivalently
\beq
      [-\oh \hbar^2  {V \over \sqrt{|VG|}} {\partial
\over \partial q^n} \sqrt{|VG|} {1\over V} G^{nm} {\partial \over \partial q^m}
 +  {1 \over 3} \hbar^2 V {\cal R} - {2m_0^2 \E \over D+1}V] \Psi = 0
\label{WD}
\eeq
This equation may look more recognizable if we
identify
\beq
M^2 = - {2 m_0^2\E \over D+1}
\eeq
Then reorder the factors in the Laplacian and drop the curvature scalar
term to obtain
\beq
     "~~ [-\oh \hbar^2  G^{nm} {\partial
\over \partial q^n} {\partial \over \partial q^m}
 +  M^2 V] \Psi  = 0~~"
\eeq
where the quotation marks mean that this equation is correct up
to operator-ordering terms.  From this it is easy to see
that the equation satisfied by stationary states, eq. \rf{WD}, is
simply the Wheeler-DeWitt constraint equation, with the dimensionful
parameter identified as $M^2=-2m_0^2 \E / (D+1)$, and with a particular
choice of operator ordering.  Once again, we see that all stationary
states satisfy a Wheeler-DeWitt equation, but with different mass
parameters.

   As in the last section, the classical equations of motion are obtained
from the classical Hamiltonian given by \rf{Hclass}:

\beq
      H_c = {D+1 \over 4m_0^2 V} G^{nm} p_n p_m
\eeq
and, setting $H_c=\E$ and $M^2 = -2m_0^2 \E /(D+1)$, we obtain the
classical constraint equation
\beq
      {\cal H}_M \equiv {1 \over 2M} G^{nm} p_n p_m - M V = 0
\label{H_M}
\eeq
Hamilton's equations give us
\bea
      {d q^n \over d \t} &=& {\partial H_c \over \partial p_n} =
{D+1 \over 2m_0^2 V} G^{nm} p_m
\non \\
      {d p_n \over d \t} &=& -{\partial H_c \over \partial q^n} =
-{D+1 \over 4m_0^2}\left[ {1\over V}
{\partial G^{ij} \over \partial q^n} p_i p_j - {1 \over V^2}
{\partial V \over \partial q^n} G^{ij} p_i p_j \right]
\non \\
    &=& -{(D+1)M \over 2m_0^2 V}\left[ {1\over 2M}
{\partial G^{ij} \over \partial q^n} p_i p_j - M
{\partial V \over \partial q^n} \right]
\eea
which can be rewritten as

\bea
       {\cal H}_M &=& 0
\non \\
       {d q^n \over d \t} &=& f(q) {\partial {\cal H}_M \over \partial p_n}
\non \\
       {d p_n \over d \t} &=& - f(q) {\partial {\cal H}_M \over \partial q^n}
\label{Hamilton}
\eea
where
\beq
         f(q) = {(D+1)M \over 2m_0^2 V(q)}
\label{f}
\eeq
and where ${\cal H}_M$, given in \rf{H_M}, is the minisuperspace
Hamiltonian ${\cal H}$ with the parameter $m_0$ replaced by $M$.

  Apart from the factor of $f(q)$, equations \rf{Hamilton}
are simply the classical equations of motion of the minisuperspace
action (with $m_0 \ra M$).  So long as V is non-zero,
$f(q)$, like the lapse function $N(t)$, is irrelevant in
determining the classical trajectory,
which depends only on the directions of the vectors $\partial_\t q^n$ and
$\partial_\t p_n$ in phase space, and not on their magnitudes.  The
magnitudes of these vectors only determine the rate (compared to some
analog of proper time in minisuperspace)  at which the evolution
parameter runs along the classical trajectory.

   This point can be made in another way.  Suppose we pick an initial
point in phase space $\{q,p\}_0$ and solve the equations \rf{Hamilton}.
Denote the solutions $\bar{q}^n(\t)$ and $\bar{p}_n(\t)$.  Then, define

\beq
         N(\t )_{\{q,p\}_0} \equiv f(\bar{q}(\t))
\label{lapse}
\eeq
It is easy to see that that $\bar{q}^n(\t)$ and $\bar{p}_n(\t)$ are a
solution of Hamilton's equations for the original minisuperspace action
\rf{Smini}
\bea
       {\cal H}_M &=& 0
\non \\
 {d q^n \over d \t} &=& N(\t )_{\{q,p\}_0} {\partial {\cal H}_M \over \partial
p^n}
\non \\
 {d p_n \over d \t} &=& - N(\t )_{\{q,p\}_0} {\partial {\cal H}_M \over
\partial q^n}
\label{Hamilton1}
\eea
with $m_0$ replaced by $M$, and with a particular choice
\rf{lapse} for the lapse function (which will be different for each
classical trajectory).

\section{Quantum Gravity}

   The action of quantum gravity, in the ADM decomposition, is

\bea
       S &=& \int d^4x \left[ p^{ij} {\partial g_{ij} \over \partial t}
- N{\cal H}_x(\k_0^2) - N_i {\cal H}^i_x \right]
\non \\
   {\cal H}_x(\k_0^2) &=& \k_0^2 G_{ijkl}p^{ij}p^{kl}
+ {1\over \k_0^2} \sqrt{g}(-R + 2 \L)
\non \\
       {\cal H}^i_x &=& -2 p^{ik}_{~~;k}
\non \\
       G_{ijkl} &=& {1 \over 2 \sqrt{g}}(g_{ik}g_{jl}+g_{il}g_{jk}
                                       - g_{ij}g_{kl})
\label{SADM}
\eea
where $\rg$ is the root determinant of the
three-metric $g_{ij}$.  At the classical level, $\k_0^2 = 16 \pi G_N$,
where $G_N$ is Newton's constant.  It will be seen that this identification
is modified at the quantum level, much as $m_0^2$ was replaced by
$M^2$ in the examples of the preceding sections.

  The presence of the shift functions $N_i$ is a serious complication, as
compared to the minisuperspace models of the preceding section, where
the shift functions were absent.  The problem is that the classical
equation of motion

\beq
   {\partial g_{ij} \over \partial t} = 2\k_0^2 NG_{ijnm}p^{nm} + N_{i;j}
+ N_{j;i}
\label{p}
\eeq
which are used to solve for the momenta in terms of $\partial_t g_{ij}$,
contain derivatives of the $N_i$.  These shift functions can be solved for,
in principle, by substituting the expression for $p_{ij}$ obtained
from \rf{p} into the supermomentum constraints ${\cal H}^i_x=0$, which gives
the $N_i$ in terms of the lapse $N$, and then
substituting these $N_i$ into the Hamiltonian constraint ${\cal H}_x=0$ to
solve
for the lapse.  However, since the $N_i$ are determined, in this way,
by complicated partial differential equations, this procedure leads to
intractable expressions, and there is no simple form for $S[g',g]$.

   A great simplification is achieved if we instead set $N_i=0$ from the
beginning.  In that case the supermomentum constraint is not obtained
by extremizing the action, and must somehow be recovered by imposing
an operator constraint on physical states

\beq
           Q_x[p_{ij},g_{ij}] \Psi = 0
\eeq
where the subscript $x$ indicates that there is a separate constraint
at each point, and of course these must be mutually
consistent, as well as consistent with the $\t$-evolution operator.
These constraints will be obtained below.  For the moment, we just
set $N_i=0$ and proceed.  Denote

\bea
        \{a=1-6\} &\leftrightarrow& \{(i,j),~i \le j \}
\non \\
             q^a(x) &\leftrightarrow& g_{ij}(x)
\non \\
             p_a(x) &\leftrightarrow&
\left\{ \begin{array}{rr}
             p^{ij}(x)~~~~~(i=j) \\
           2 p^{ij}(x)~~~~~(i<j) \\
        \end{array} \right.
\non \\
             G_{ab}(x) &\leftrightarrow& G^{ijnm}(x)
\eea
in order to compress the number of indices somewhat. Hamilton's
equations with $N_i=0$ give us

\beq
      p_a = {1 \over 2 \k_0^2 N} G_{ab} \q^b
\eeq
Inserting this into the constraint equation
\beq
       0 = {\cal H}_x = {1 \over 4\k_0^2 N^2}G_{ab}\q^a \q^b
+ {1\over \k_0^2}\rg U
\eeq
where
\beq
       U \equiv  -R + 2 \L
\eeq
and solving for the lapse, gives
\beq
         N = \left[-{1 \over 4 \rg U}G_{ab}\q^a \q^b \right]^{1/2}
\eeq
so we have
\bea
       \D S = S[q',q] &=&-{1 \over \k_0^2}\int d^3x \int^{\D t}_0 dt
\sqrt{-g^{\oh} U G_{ab} \q^a \q^b}
\non \\
         &=& -{1 \over \k_0^2} \int d^3x (\sqrt{-(g^{\oh} U G_{ab})_0
\D q^a \D q^b}
                   + O(\D q^2)
\non \\
         &=& -{1 \over \k_0^2}\int d^3x (\rg)_0 (\sqrt{-(\G_{ab})_0
\D q^a \D q^b} + O(\D q^2)
\label{DS}
\eea
where $\D q^a = q^a-q'^a$, and we define
\beq
       \G_{ab} \equiv {1 \over \rg} U G_{ab}
\eeq
The notation $(..)_0$ means that the quantity in
parenthesis is to be evaluated at $\D q =0$; i.e. at $q'$.

   Once again, we implicitly
rotate the signature of the superspace metric $G_{ab}$
to Euclidean signature, and extract a factor of $i$, before
evaluating the transfer matrix.  Note that it
is the signature of superspace, not the signature of spacetime, which
is rotated to the Euclidean value.  At the end,
of course, we rotate back to the usual signature of superspace, according
to the prescription \rf{Hlim}. It was found, in the minisuperspace
example of the previous section, that the evolution operator is
Hermitian in an appropriate measure $\m(q)$, despite the fact that
the potential term $V$ was not positive definite.
It is expected, for similar reasons, that hermiticity of $H$ will also be
maintained in full quantum gravity, where $U$ is not positive definite.
\footnote{An actual proof of hermiticity, however, would involve the
regularization/operator-ordering issues discussed below.}

   The transfer matrix is obtained from

\bea
     \psi(q',\t+\e) &=& \int Dq \m(q) e^{-\D S/\reh } \left[ \psi(q')
+ \int d^3x \left({\d \psi \over \d q^a(x)}\right) \D q^a(x)
\right.
\non \\
    & & \left. + \oh \int d^3x d^3y \left( {\d^2 \psi \over \d q^a(x)
\d q^b(y)}\right) \D q^a(x) \D q^b(y) + ... \right]
\non \\
    &=& 1 + \e [T_0 + T_1 + T_2] + O(\e^2)
\label{Tn}
\eea
where the $T_n$ represent terms with $n$ derivatives of $\psi$.  To
find these terms, we need to evaluate
\bea
    <\D q^a(x_1) \D q^b(x_2)> &=& \int D(\D q) \; (\m)_0
\D q^a(x_1) \D q^b(x_2)
\non \\
& & \times \exp\left[-{1 \over \k_0^2}\int d^3x (\rg)_0 (\sqrt{(\G_{ab})_0
\D q^a \D q^b}/\reh \right]
\eea
This quantity is highly singular, and, at this point, regularization is
unavoidable.

    Unfortunately, the problem of regularizing a functional integral
\linebreak non-perturbatively, in such a way as to preserve diffeomorphism
invariance, is unsolved.  So I will have to resort to a naive regulator,
and replace the continuum degrees of freedom, labeled by $x$, by a
finite set, labeled $\{n\}$, which are taken to represent regions
of equal volume.  We make the correspondences

\bea
        \D q^a(x) &\leftrightarrow& \D q^a(n)
\non \\
        \int d^3x \rg &\leftrightarrow& {\V \over N_p} \sum_{n=1}^{N_p}
\non \\
         {\d \over \d q^a(x)}  &\leftrightarrow& {N_p \sqrt{g(n)} \over \V}
{\partial \over \partial q^a(n)}
\non \\
          Dq &\leftrightarrow& \prod_n d^Dq(n)
\label{discrete}
\eea
where $\V$ is the three-volume. For pure quantum gravity, there are
six independent components of $g_{ij}$, so D=6.
Obviously, with this naive discretization we lose
diffeomorphism invariance of the integration measure, and we can
expect to make mistakes, even after taking the continuum limit,
on certain expressions which depend crucially on
the invariance of the measure.  In the transfer matrix formulation,
it is the operator ordering, and the presence of terms analogous
to $R$ in eq. \rf{Hcurv} (or ${\cal R}$ in \rf{WD}),
which are sensitive to the measure.
Operator-ordering problems are notorious in quantum gravity, and
I will not try to solve them here.  From here on, only the principal
term of the evolution operator (coming from $T_2$ in eq. \rf{Tn}),
will be determined explicitly.  This term is relatively insensitive
to the measure, and is the only term which is
important for the classical limit.
But even for the $T_2$ term, the ultimate justification for the
prescription \rf{discrete} above will be a posteriori.

   With the above caveats duly noted, we find
\bea
    <\D q^a(n) \D q^b(m)> &=& \int \prod_k d^Dq(k)  \; (\m)_0 \D q^a(n)
\D q^b(m)
\non \\
  & & \exp\left[-{\V \over N_p \k_0^2} \sum_k
\sqrt{\G_{ab})_0 \D q^a(k) \D q^b(k)}/\reh \right]
\non \\
       &=& \d_{nm}(D+1) \hbar \e {\k_0^4 N_p^2 \over \V^2} \G^{ab}
\eea
then
\bea
      \e T_2 &=& \oh \sum_n \sum_m {\partial^2 \psi \over \partial
q^a(n) \partial q^b(m)} <\D q^a(n) \D q^b(m)>
\non \\
  &=& \oh (D+1) \e \hbar {\k_0^4 N_p^2 \over \V^2} \sum_n
{\partial^2 \psi \over \partial q^a(n) \partial q^b(n)} \G^{ab}
\non \\
  &\ra& \e \hbar {\b \k_0^4 \over \V} \int d^3x \; U^{-1} G^{ab}
{\partial^2 \psi \over \partial q^a \partial q^b}
\eea
where $\b$ is a divergent, dimensionless constant
\beq
      \b \equiv \oh (D+1) N_p
\eeq
proportional to the number of degrees of freedom.  Finally, from
$T_2$ we get

\beq
      H = -\hbar^2 {\b \k_0^4 \over \V} \int d^3x \; U^{-1} G^{ab}
{\d^2 \over \d q^a \d q^b}
\label{Hgrav}
\eeq
where, again, operator-ordering/measure terms have been dropped.

    However, this evolution operator does not yet give
us general relativity, not even in the classical limit.  This is
easy to see from, e.g., the WKB approach, since the Hamilton-Jacobi
equation corresponding to \rf{Hgrav} is
\beq
      \E = {\b \k_0^4 \over \V} \int d^3x \; U^{-1} G^{ab}
{\d S \over \d q^a} {\d S \over \d q^b}
\eeq
which is not the Einstein-Hamilton-Jacobi equation of gravitation.  Now,
although the transfer matrix formulation as presented in eq. \rf{tmat}
and \rf{measure} was guaranteed to recover the correct classical limit,
it is not hard to see what went wrong in this case.  The difficulty,
as already mentioned above, comes from setting the shift functions
$N_i=0$ at the beginning; this means that the supermomentum
constraints ${\cal H}^i_x=0$ were lost from the start.  One could try
to simply reintroduce these as physical state constraints
${\cal H}^i_x \Psi = 0$, but this still doesn't
lead to the Einstein-Hamilton-Jacobi equation.

   Note however, that because of the factor $1/\V$ in the $\t$-evolution
operator $H$ of eq. \rf{Hgrav}, the Schrodinger equation can be
written in the form
\beq
      \int d^3x Q_x \Psi = 0
\label{QS_eq}
\eeq
where
\beq
       Q_x = -\hbar^2 \b \k_0^4 U^{-1} G^{ab} {\d^2 \over \d q^a \d q^b}
 -i \hbar \rg {\partial \over \partial \t}
\eeq
The extra constraints which need to be imposed on the physical states,
which then generate the usual constraint algebra of general relativity,
are simply
\beq
        Q_x \Psi = 0
\label{Qconstraint}
\eeq
at every point $x$.  It is not hard to see why.  Consider an arbitrary
solution of the Schrodinger equation
\beq
       \Psi(q,\t) = \sum_{\E} a_\E e^{-i\E \t/\hbar} \Phi_\E(q)
\eeq
Since the $a_\E$ are arbitrary, the Q-constraint \rf{Qconstraint}
requires that for each stationary state
\beq
\left\{ -\hbar^2  G^{ab} {\d^2 \over \d q^a \d q^b}
+ \rg \left({-\E \over \b \k_0^4}\right) U \right\}
\Phi_{\E} = 0
\label{WD2}
\eeq
Identifying
\bea
          \k^2 = 16 \pi G_N &=& \sqrt{-{\b \over \E}} \k_0^2
\non \\
   M_P^2 &=& \sqrt{-{\E \over \b}} {16 \pi \hbar \over \k_0^2}
\eea
as Newton's constant and the Planck mass, respectively,
we see that eq. \rf{WD2} is just the Wheeler-DeWitt equation
\beq
        {\cal H}_x(\k^2) \Phi_\E =
\left\{ -\hbar^2 \k^2  G^{ab} {\d^2 \over \d q^a \d q^b}
+ {1 \over \k^2} \rg U \right\}
\Phi_{\E} = 0
\label{WD3}
\eeq
with Newton's constant inversely proportional to $\sqrt{-\E}$.

   Finally, we invoke the Moncrief-Teitelboim interconnection theorem
\cite{Moncrief}, which says that if a state satisfies the (Wheeler-DeWitt)
Hamiltonian constraint \rf{WD3} at every point $x$, then that state
also satisfies the supermomentum constaints
\beq
        {\cal H}^i_x \Phi_\E = 0
\label{Hi}
\eeq
at every point $x$.  In this way, the supermomentum constraints that
were lost at the outset by fixing $N_i=0$ have been recovered.  Further,
given that the Hamiltonian and supermomentum constraints are consistent
(commutators close on the Poincare algebra), and that the $Q_x$ constraints
\rf{Qconstraint} are implied by the Hamiltonian constraints \rf{WD3},
it follows that the $Q_x$ constraints are consistent not only with the
evolution operator $H$, but also with each other.

   The Einstein-Hamilton-Jacobi equations follow directly from
a WKB treatment of the Wheeler-DeWitt constraints, and
the classical limit, obtained by replacing $-i \hbar \d / \d q^a$
with $p_a$, follows in complete analogy to the minisuperspace case.
Since the proper constraint algebra has been obtained, it is
fairly obvious that that the correct classical equations must fall out.
But it is still nice to see this explicitly.  We begin with the
classical limit of the evolution operator

\bea
       H_c &=& {1 \over \V} \int d^3x \; \b \k_0^4 U^{-1}G^{ab}p_a p_b
\non \\
           &\equiv& {1 \over \V} \int d^3x \; H_{cx}
\label{Hcx}
\eea
Begining from a set of initial data $\{q,p\}_0=\{g_{ij},p^{ij}\}_0$
consistent with the supermomentum constraints
\beq
          {\cal H}^i[\{q,p\}_0] = 0
\eeq
we have
\beq
           \E \equiv H_c[\{q,p\}_0]
\label{C1}
\eeq
and consistency of the initial data with the $Q_x=0$ constraints
gives us also
\beq
            H_{cx} = \rg \E
\label{C2}
\eeq
Then Hamilton's equations, derived from $H_c$, are
\bea
     {dq^a(x) \over d \t} &=& {\b \k_0^4 \over U \V} 2 G^{ab} p_b
\non \\
   &=& \int d^3x' {\b \k_0^4 \over U \V \k^2}
{\d \over \d p_a(x)} {\cal H}_{x'}(\k^2)
\eea
and
\bea
 {dp_a(x) \over d \t} &=& - \left\{ -{1 \over \V^2} {\d \V \over
 \d q^a(x)} \int d^3x' H_{cx'} - {1 \over \V} \int d^3x'
{1 \over U^2} {\d U \over \d q^a(x)} \b \k_0^4 G^{bc}p_b p_c \right.
\non \\
& & \left. + {1 \over V} \int d^3x' {\b \k_0^4 \over U}
{\d \over \d q^a(x)} G^{bc} p_b p_c \right\}
\eea
Applying eq. \rf{C2}
\bea
  {dp_a(x) \over d \t}
  &=& -\int d^3x' {1 \over U \V} \left\{ -{\d \rg \over \d q^a(x)}
U \E - {\d U \over \d q^a(x)} \rg \E \right.
\non \\
  & & \left. + \b \k_0^4 {\d \over \d q^a(x)} G^{bc}p_b p_c \right\}
\non \\
  &=& - \int d^3x' {\b \k_0^4 \over U \V \k^2} {\d \over \d q^a(x)}
\left[ \k^2 G^{bc}p_b p_c + {1\over \k^2} \rg U \right]
\non \\
  &=& - \int d^3x' {\b \k_0^4 \over U \V \k^2}{\d \over \d q^a(x)}
{\cal H}_{x'}(\k^2)
\eea
The evolution parameter $\t$ has units of mass$\times$length.  To give
it conventional units, rescale $\t \ra t=\t/M_P$, and define
\beq
      F_x[q] \equiv {M_P \b \k_0^4 \over U(x) \V \k^2}
\eeq

Then the full set of gravitational field equations derived from
$H_c$ plus the constraints is
\bea
   {d q^a(x) \over d t} = M_P{\d H_c \over \d p_a(x)}
&=& \int d^3x' F_{x'}[q]{\d \over \d p_a(x)} {\cal H}_{x'}(\k^2)
\non \\
   {d p_a \over d t} = - M_P{\d H_c \over \d q^a(x)}
&=& - \int d^3x' F_{x'}[q]{\d \over \d q^a(x)} {\cal H}_{x'}(\k^2)
\non \\
            {\cal H}_x(\k^2) &=& 0
\non \\
            {\cal H}^i_x &=& 0
\label{gravity}
\eea
To see that this is classical Einstein gravity, let
$\bar{q}^a(x^i,t)$ and $\bar{p}_a(x^i,t)$ be a solution of \rf{gravity}
for some set of initial data $\{q,p\}_0=\{g_{ij},p^{ij}\}_0$
compatible with the constraints.  Then, as in the minisuperspace case,
define
\beq
          N(x,t)_{\{p,q\}_0} \equiv F_x[\bar{q}(t)]
\label{lapse1}
\eeq
and one sees that $\bar{q}^a(x^i,t)$ and $\bar{p}_a(x^i,t)$ is a
solution of
\bea
   {d q^a(x) \over d t} &=& {\d H_{Einstein} \over \d p_a(x)}
\non \\
   {d p_a(x) \over d t} &=& -{\d H_{Einstein} \over \d q^a(x)}
\non \\
    H_{Einstein} &=& \int d^3x \; N(x,t)_{\{p,q\}_0} {\cal H}_x(\k^2)
\non \\
            {\cal H}_x(\k^2) &=& 0
\non \\
            {\cal H}^i_x &=& 0
\label{Einstein}
\eea
which are simply the Einstein field equations in Hamiltonian form,
with lapse \rf{lapse1}, which is in general different for each classical
solution, and shift $N_i=0$.

\section{Discussion}

   To summarize: given an initial "wavefunction of the Universe"
$\Psi[q,\t_0]$, it has been shown how to trace its subsequent evolution
along a trajectory in the Hilbert space of physical states, with the
trajectory parametrized by $\t$.  The $\t$-evolution equation for
quantum gravity, up to factor-ordering terms, is
\bea
  i\hbar \partial_\t \Psi[q,\t] &=& H \Psi[q,\t]
\non \\
   &=& -\hbar^2 {\b \k_0^4 \over \V} \int d^3x U^{-1} G^{ab}
{\d^2 \over \d q^a \d q^b} \Psi[q,\t]
\label{Seqgrav}
\eea
where the space of physical states is spanned by the eigenstates
\beq
              H \Phi_\E [q] = \E \Phi_\E [q]
\eeq
each of which satisfies the constraint algebra
\bea
            {\cal H}_x(\k^2) \Phi_\E &=& 0
\non \\
            {\cal H}^i_x \Phi_\E &=& 0
\non \\
            \k^4 &=& -{\b \over \E} \k_0^4
\label{algebra}
\eea
and where ${\cal H}_x(\k^2)$ denotes the superHamiltonian constraint
operator, with Newton's constant $G_N=\k^2/16\pi$.
Modulo operator-ordering
issues, {\it every stationary  physical state in the transfer
matrix formulation satisfies the usual constraint algebra of general
relativity, but each with a different value of Newton's constant, where
$G_N^2$ is inversely proportional to
the eigenvalue $\E$ of the evolution operator}. The number of degrees of
freedom, in this formulation, is therefore $2 \times \infty^3 + 1$; i.e.
two degrees of freedom per point, which is the degrees of
freedom of states satisfying the constraint algebra \rf{algebra},
and one extra degree of freedom corresponding to the Planck mass.
This is only one more degree of freedom (overall, not per point)
than in the standard formulations; there should be no danger of, e.g.,
the graviton aquiring a mass.

   Thus, in the transfer-matrix formulation, time-evolution of states is
recovered at a modest price:  the Planck mass (inverse Newton's
constant) becomes a dynamical quantity, analogous to energy in
non-relativistic quantum mechanics.  As is the case for the time
parameter in non-relativistic quantum mechanics,
or in quantum field theory on a fixed
spacetime background, the evolution parameter $\t$ is {\it only} an
evolution parameter; it is not an observable, and there is no
operator acting on the Hilbert space which corresponds to $\t$.  This
avoids the problems encountered in relativistic quantum mechanics,
as well in standard formulations of canonical quantum gravity, where
the approach is to identify one of the operators in the theory as an
evolution parameter.

   We note that the spectrum of the evolution operator $H$ in
\rf{Seqgrav} is unbounded from below; this is true of all the
systems considered in this paper.  For closed systems (such as the
Universe) which do not interact with anything external, this absence
of a ground state is not a problem.  The eigenvalue $\E$ is a constant
of motion, and its distribution cannot change.

    In non-relativistic quantum mechanics, any non-stationary state has
a certain dispersion in its energy.  For quantum gravity, the
corresponding statement is that since the Universe is non-stationary,
there must be a certain dispersion in $\E$, the eigenvalue of the
evolution operator.  This implies dispersion
in the Planck mass $M_P$, or, equivalently, Newton's constant.  Since
$M_P^4 \propto -\E$ is conserved by the evolution equation,
the fractional dispersion of Newton's constant $\D G_N / G_N$
should be a dimensionless constant of nature.  Depending on how large
this dispersion is, there could conceivably be observational consequences.
I hope to return to this question at a later time.

\vspace{33pt}

\noindent {\Large \bf Acknowledgements}{\vspace{11pt}}

   I am grateful to Noboru Kawamoto for a discussion which
inspired this study.  I am also happy to thank Jan Ambjorn,
John Moffet, and Holger Nielsen for helpful discussions, and
to acknowledge support from the Danish Research Council.


\end{document}